\documentclass[11pt,a4paper]{article}
\usepackage{jcappub}
\usepackage{graphicx}

\def\apj{{\it Astrophys.~J.}}
\def\apjs{{\it Astrophys.~J. Suppl.}}

\def\prd{{\it Phys.~Rev.~D}}
\def\prl{{\it Phys.~Rev.~Lett.}}

\def\mnras{{\it Mon.~Not. Roy.~Astr.~Soc.}}
\def\ijmpd{{\it Int.~J.~ Mod. Phys. D}}

\def\AnA{{\it Astron. Astrophys.}}
\def\grg{{\it Gen. Rel. Grav.}}
\def\ARAnA{{\it Ann. Rev. Astron. Astrophys.}}

\title{Observational Cosmology and the Cosmic Distance Duality Relation}
\author[a]{Remya Nair,}
\author[a]{Sanjay Jhingan}
\author[b]{and Deepak Jain}

\affiliation[a]{Centre for Theoretical Physics,\\ Jamia Millia
Islamia, New Delhi 110025, India}

\affiliation[b]{Deen Dayal Upadhyaya College, \\
University of Delhi, New Delhi 110015, India}

\emailAdd{remya$_{-}$phy@yahoo.com}
\emailAdd{sanjay.jhingan@gmail.com} \emailAdd{djain@ddu.du.ac.in}

\abstract{We study the validity of cosmic distance duality
relation between angular diameter and luminosity distances. To
test this duality relation we use the latest Union2 Supernovae Type
Ia (SNe Ia) data for estimating the luminosity distance. The estimation
of angular diameter distance comes from the samples of galaxy
clusters (real and mock) and FRIIb radio galaxies. We parameterize the distance
duality relation as a function of redshift in six different ways.
Our results rule out some of the parameterizations significantly.}

\keywords{supernovae type Ia - standard candles, galaxy clusters, dark energy theory
}

\begin{document}

\maketitle

\section{Introduction}
To understand the physics behind the late time cosmic acceleration
is one of the most challenging problem in the cosmology. Many
observational methods using distant type Ia supernovae (SNe Ia), Cosmic Microwave
Background (CMB), Galaxy Clusters, Baryon Acoustic Oscillations
(BAO) etc. have been used to probe the mechanism behind this
positive acceleration \cite{fri}. These tools are based on the
fundamental assumption that total number of photons are conserved on
cosmic scales. This is one of the key assumptions in the
relationship between the luminosity distance, $d_L$, and the angular
diameter distance, $d_A$,  \cite{bk},
i.e.

\begin{equation}\label{ddual}
\frac{d_L}{d_A(1+z)^2} \, = \,1 \;.
\end{equation}
This equation is known as Distance Duality (DD) relation \cite{dd}.
It is natural to check the validity of this equation which may play
a key role in observational cosmology. In particular it plays a
vital role in   galaxy observations, CMB observations and
gravitational lensing \cite{ellisgrg}. Furthermore, violation in the
DD relation may point to a failure of the metric theory of gravity
in explaining the background dynamics of the universe and emergence
of new physics. Whether this DD relation can shed some light on the
presence of exotic physics or not was first explored by Bassett $\&$
Kunz (2004) {\cite{bk}}. They ruled out non-accelerating models of
universe (replenishing dust model) by more than $ 4 - \sigma$ level.
In this context to study the validity of DD relation (\ref{ddual}),
we analyse the following red-shift dependence of DD
\begin{equation}\label{ddrelation}
\eta(z) \equiv \frac{{d_L}}{d_A(1+z)^2 } \;.
\end{equation}

Uzan et al.(2004) used the combined measurements of
Sunyaev-Zeldovich effect and  X-ray emission data of galaxy
clusters to study the violation in the DD relation \cite{uz}. They
showed that if this relation does not hold then the angular diameter
distance measured from the clusters is $d_A^{cluster}(z)= d_A\,
\eta^2$ , and hence the DD relation (\ref{ddrelation}) gets modified to
\begin{equation}\label{erro}
\eta(z) \equiv \frac{{d_A^{cluster}(1+z)^2}}{d_L}  \,.
\end{equation}
They found no significant violation of DD relation as they obtained
the value of $\eta$ close to 1. Their data set consisted of 18
galaxy clusters. Further, De Bernardis et al.(2006) found no
departure from DD relation i.e. $\eta =1$ at $1 \sigma$ C.
L.\cite{b}. They used bigger sample of 38 galaxy clusters in the
redshift range of $0.14 < z < 0.89$. Later on  Corasaniti
(2006) modeled the intergalactic dust in terms of the star formation
history of the universe and forecasted a deviation in the DD
relation due to the presence of the cosmic dust extinction
\cite{co}.

Recently Holanda et al.(2011) used two different data sets of galaxy
clusters to check the consistency of DD relation \cite{h0}. These
two data sets were based on two different geometrical descriptions
of clusters. By assuming two parameterizations for $\eta(z)$, they
showed that the data set which is based on the elliptical $\beta $
model of clusters gives better fit with DD relation as compared to
the data set based on spherical $\beta$ model. Furthermore, Holanda
et al. (2010) checked the validity of DD relation using the galaxy
cluster sample and SNe Ia  with different parameterizations  of
$\eta(z)$ \cite{h1}. Hence they obtained bounds on the parameters of
$\eta(z)$. They concluded that the best fit values of the parameters
of the $\eta(z)$ parametrization obtained through the data set based
on spherical $\beta$ model of cluster are not consistent with the DD
relation.

Lazkoz et al. found no evidence of violation in this relation at $ 2
\sigma$ level \cite{la}. They used SNe Ia  and CMB + BAO as standard
candles and standard rulers respectively to test the validity of
this relation.

Further Avgoustidis et al. demonstrated that the DD
relationship can be used to put constraints on the cosmic
transparency \cite{av}. They obtained the distances by using SNe Ia
data, BAO and measurements of the Hubble parameter, $H(z)$.
Following the same line of thought More et al. also explored the
cosmic transparency (conservation of photons phase-space density) by
using distance\textsl{} measures \cite{mo}.

More recently Lampeitl et al. analyzed  the DD relation and found no
proof of violation at the one sigma level. For this analysis, they
used SDSS II SNe data and  local BAO measurements at redshift  z= 0.2
and 0.35 \cite{lam}.

The aim of this paper is to reanalyze the validity of DD relation in
a more comprehensive manner by using  different data samples and
parameterizations. In this work we have used three different
data-sets of galaxy clusters, radio galaxies and a Mock data set to
determine angular diameter distance, along with Union2 sample of SNe
Ia. We have analyzed the validity of DD relation by using six
different parameterizations of $\eta(z)$.

The plan of the paper is as follows. In the next section we
describe various parameterizations considered in this work. The
details of the data and methodology used is also explained in this
section. We conclude with a section on results and discussion.

\section{Parameterizations, data and method}

\subsection{$\eta$(z) parameterizations}
In this work we parameterized the $\eta(z)$ with one index and two
index parameterizations. These relations are inspired by model
independent parameterizations for the dark energy equation of state
\cite{jo}. We are parameterizing $\eta(z)$ whose value stays one when
photon number is conserved, gravity is described by a
metric theory and photons travel on null geodesics. Any
significant violation from the DD relation will hint at the emergence
of new physics.
Since all these assumptions are reasonably well tested \cite{bk},
we expect $\eta$ to stay close to unity. To model any departure from
unity we parameterize $\eta$ with six parametric representations for a
possible redshift dependence of the distance duality relation:

\begin{itemize}

\item {Two index parameterizations}
\begin{align}
\eta_{I}(z) &=\eta_{0}+\eta_{1} z \\
\eta_{II}(z) &=\eta_{2} + \eta_{3} \; \frac{ z}{1+z} \\
\eta_{III}(z)&=\eta_{4} + \eta_{5}\; \frac{ z}{(1+z)^2} \\
\eta_{IV}(z)&=\eta_{6}-\eta_{7}\, \ln(1+z)
\end{align}

\item {One index parameterizations}

\begin{align}
\eta_{V}(z) &=\eta_{8} \;\left(\frac{1}{ 1+z}\right) \\
{\eta_{VI}}(z) &=\eta_{9} \; \left(\frac{1}{ 1+z}\right)
\exp\left(\frac{z}{1+z}\right)
\end{align}

\end{itemize}

Both, one and two index parameterizations have their limitations and
advantages. The two index parameterizations are more flexible in
comparison to one
index parameterizations which may completely dominate over data.

\subsection {Data}
We have used three different data sets to calculate the angular
diameter distance. For the luminosity distances, we use latest
Union2 SNe Ia data \cite{am}.  Now we consider that pair of galaxy
cluster/radio galaxy and SNe for which $\Delta z < 0.005$ \cite{h1}.
Because of this condition, the number of data points are limited to
24, 222 and 12 for the data set I, II and III respectively. If there are
more than one supernova satisfying $\Delta z < 0.005$ for a given
cluster/radio
galaxy we choose the nearest SNe to the given cluster/radio
galaxy. In case of multiple SNe at same redshift value for a
given cluster/radio
galaxy we choose that SNe which has less error bars.

\begin{itemize}

\item Data Set I: This sample consists of 25
galaxy clusters (isothermal elliptical $\beta$ model, with concordance model
for the cosmological distance-redshift relationship) \cite{fi}. By
combining together the Sunyaev-Zeldovich temperature decrements and
X-ray surface brightness observations, one can obtain the angular
diameter distance for clusters. The redshift interval for clusters
in  this sample is $0.02 < z < 0.78$. After applying the selection
criterion, the number of data points reduce to 24 in this sample.

\item Data Set II:  This bigger data set contains 578  mock values of
angular diameter distances of mock clusters, and for 222 mock
clusters we have SNe Ia
satisfying $\Delta z < 0.005$. This catalog is created by assuming
fiducial cosmology from the $WMAP7$ + $BAO$ + $H_0$ results
\cite{larson}. This data set assumes spherical isothermal $\beta$ model
for Intra-cluster medium (ICM). This mock catalog is created for overlap of
ACT/SPT + eROSITA with $25\%$ Gaussian scatter error which are
similar to the recent measurements of angular diameter distance
\cite{reese}. The estimate of angular diameter distances strongly
depends on the ICM model. The redshift distribution of this sample
is from $0.05 < z < 0.76$. For more details see the ref.'s
\cite{sa1,sa2}.

\item Data Set III: In this sample the calculation of angular diameter
distance is obtained by using the physical size of extended radio
galaxies \cite{ru}. This data set contains 20 radio galaxies up to
redshift z = 1.8 . For 12 galaxies we could find corresponding SNe Ia
where the condition $\Delta z < 0.005$ is satisfied.

\end{itemize}

\subsection{Method}
We perform the $\chi^2$ analysis to fit the parameters of the
assumed parameterizations.
\begin{equation}
\chi^2(p) = \sum_{i}\frac{({\eta_{th}(z_i,p)} -
{\eta_{obs}(z_i))^2}} { {\sigma(z_i)}^2 } \; .
\end{equation}
Where $\eta_{th}$ is the assumed form of parameterizations given by
eq.(2.1) to eq.(2.6) and $\eta_{obs}$ is the observed value of
$\eta$, which is calculated by using $d_L$ and $d_A$ at a particular value
of redshift. The
unknown parameters in the parameterizations are denoted by the
variable $p$. To draw the likelihood contours at 1, 2  and 3$\sigma$,
$\Delta \chi^2 = \chi^2 -\chi^2_{min} =$ 2.3, 6.17 and 11.8
respectively in two
dimensional parametric space. For one parameter fitting, $\Delta
\chi^2 = $ 1, 4 and 9 for 1, 2 and 3
$\sigma$ C.L. respectively.

In order to find $\eta_{obs}$  for cluster, we need the observed
value of $d_A^{cluster}$ and $d_L$ as mentioned in eq. (\ref{erro}).
The luminosity distance is obtained by using the latest Union2 SNe
Ia data. The angular diameter distance, $d_A^{cluster}$, is obtained
directly by using the data set I and II. While calculating the error
bars on $\eta_{obs}$, we only considered the errors in the angular
diameter distance values. The errors in the $d_{L}$ values are quite
small compared to the errors in $d_{A}$, and hence are neglected
\cite{h1}. We know that the variance of a dependent variable
$f(x_{j})$, that depends on, say, n number of independent variables
$x_{j}$, is given by the error propagation equation as
\begin{equation}
\sigma_{f}^{2} = \sum_j\left[\sigma_{x_{j}}^{2} \left(\frac{\partial
f}{\partial x_{j}}\right)^2\right] \; .
\end{equation}
Hence, the error in $\eta_{obs}$ for galaxy
 clusters (mock and real) data set is
\begin{equation}
\sigma_{\eta_{obs}(i)} = \sigma_{d_{A}(i)} \frac{(1 + z_{i})^2}{d_{L}(i)} \; .
\end{equation}
Similarly the $\eta_{obs}$
for the radio galaxies can be obtained by assuming  $d_A^{radio} =
d_A$ (R. A. Daly, private communication) in eq. (\ref{ddrelation}). The
angular diameter distance for radio galaxies  is given in terms of the
dimensionless coordinate
distance, $y$, as
\begin{equation}
d_{A}(i) = \frac{y(z_i)}{H_0 (1 + z_i)} \; ,
\end{equation}
where  $y = a_{0} r H_0 $, and $a_0 r$ is the coordinate distance
\cite{ru}. Treating Hubble constant, $H_0$, as a nuisance parameter
we marginalize over it by assuming gaussian prior,  $H_0 = 74.2 \pm
3.6 \; km s^{-1} Mpc^{-1}$ \cite{re}. Finally, the uncertainty on
the calculated  value of $\eta_{obs},$ for radio galaxies can be
obtained by using the formula
\begin{equation}
\sigma_{\eta_{obs}(i)} = \sigma_{d_{A}(i)} \frac{d_{L}(i)}{[d_{A}(i)
\;(1  + z_{i})]^2} \;.
\end{equation}
In this analysis distances are obtained by assuming the flat
$\Lambda$ CDM universe.

\section{Results and Discussions}

In this work, we study the validity of DD relation using different data sets. We assume six general
parameterizations of $\eta(z)$ completely in analogy with the
varying equation of state of dark energy, $\omega(z)$. In these
different parameterizations it is assumed that $\eta(z)$ evolves with
redshift.

Our results are summarized as follows:

\begin{enumerate}

\item We find the best fit values of the parameters in all the six
parameterizations with data set I.  This data set contains angular
diameter of 24 galaxy clusters. Using $\chi^2$ minimization
technique we obtain the best fit values of the parameters within
$1\sigma$ range shown in Table \ref{tab:datasetI}.

\begin{table}[ht]
  \caption{Best fit values for all parameterizations - data set I}
  \vspace{0.2cm}
  \label{tab:datasetI}
  \centering
  \begin{tabular}{c c c}
    \hline
  { $\chi^2_{\nu}$} & {Parameters} & {Parameters} \\
    \hline
1.217 \quad & $\eta_{0} = 0.999 \pm 0.144$ \quad&
 $\eta_{1}=
 -0.058 \pm 0.507$ \quad\\
1.216 \quad&  $\eta_{2} = 1.007 \pm 0.170$ \quad&
 $\eta_{3} = -0.118 \pm 0.822 $
 \quad\\
1.216 \quad& $\eta_{4} = 1.013 \pm 0.205$ \quad&
 $\eta_{5} =
 -0.198 \pm 1.345$ \quad  \\
1.216 & $\eta_{6} = 1.003 \pm 0.157$ \quad& $\eta_{7} =   ~~0.087 \pm 0.651$ \\
1.337 & $\eta_{8} = 1.194 \pm 0.061$ & \\
1.169 &  $\eta_{9} = 1.01~ \pm 0.051$ & \\\hline
    \hline
  \end{tabular}
\end{table}

\begin{figure}[ht]
\centering \framebox{
\includegraphics[width=10cm, angle=0]{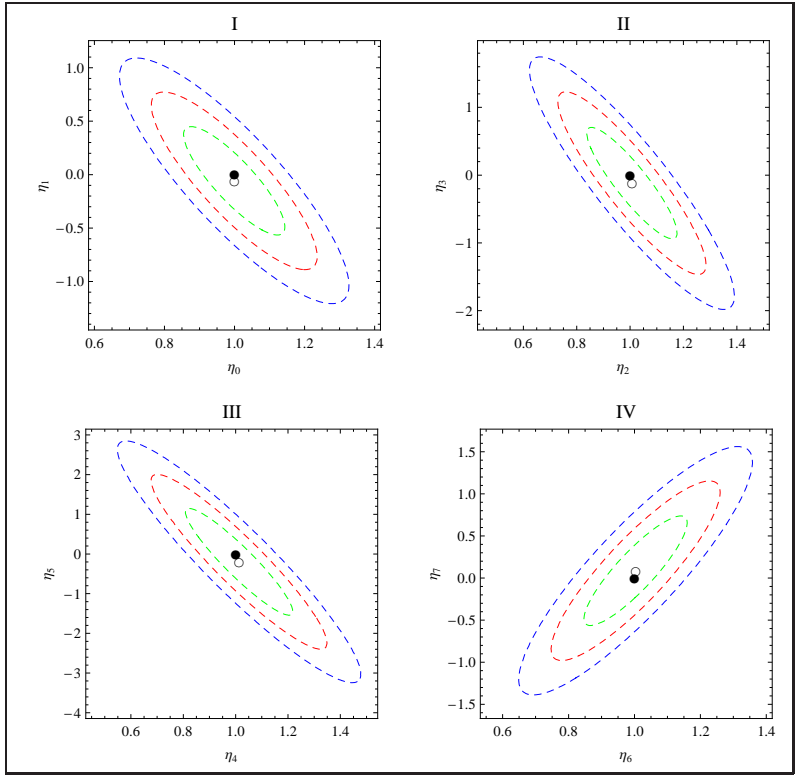}}
\caption{ $1 \sigma, 2\sigma$ and $3\sigma$ contours in $\eta_i -
\eta_j$ plane with data set I. The
  position of filled circle in the contours indicate the point where
  $\eta(z= 0) =
  1$. The position of empty circle indicates the best fit value of the
  parameters.
  }\label{filip2pa}
\end{figure}

Here $\chi^2_{\nu}$ is  reduced-$\chi^2$, or $\chi^2$ per degree of
freedom. All the two index parameterizations do support DD relation
with in 1 $\sigma$ C.L. as shown in Fig. \ref{filip2pa} for data set
I.  In one index parameterizations, $\eta_{V}(z)$, shows significant
deviation from DD relation, where as $\eta_{VI}(z)$ is in agreement
with the DD relation (see Fig. \ref{filip1p}).

\begin{figure}[ht]
\centering \framebox{
\includegraphics[width=8cm, angle =0]{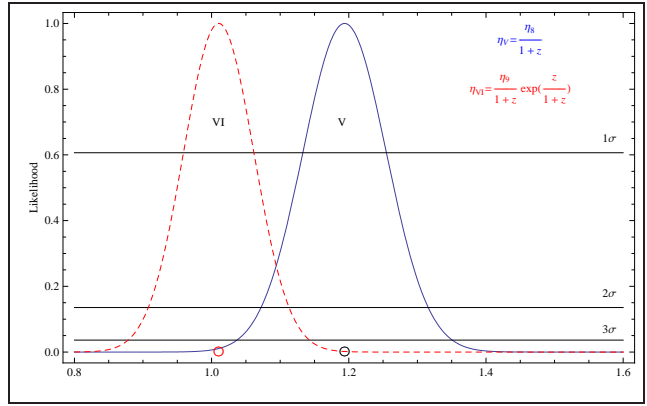}}
\caption{ Likelihood distribution function from the data Set I. The
  position of empty circle indicates the best fit value of the
  parameters
}\label{filip1p}
\end{figure}

\item  The data set II which contains the mock catalog of angular
diameter distances of galaxy clusters is  the biggest data used for
this purpose so far. It consists of 222 galaxy clusters after using the
selection criterion ($\Delta z < 0.005$). The best fit values for
all the parameterizations with in $1\sigma$ are shown in Table
\ref{tab:datasetII}.

\begin{table}[ht]
  \caption{Best fit values for all parameterizations - data set
  II } \vspace{0.2cm}
  \label{tab:datasetII}
  \centering
  \begin{tabular}{c c c}
    \hline
   {\bf $\chi^2_{\nu}$} & {Parameters} & {Parameters} \\
    \hline
  1.076 & $\eta_{0} = 0.973 \pm 0.048$ & $\eta_{1} =  -0.108
\pm 0.159$ \\
1.078 & $\eta_{2} = 0.971 \pm 0.058$ & $\eta_{3} = -0.138
\pm 0.267$ \\
1.080 & $\eta_{4} = 0.972 \pm 0.072 $ & $\eta_{5} =
-0.141 \pm 0.453$ \\
1.077 & $\eta_{6} = 0.972 \pm 0.053$ & $\eta_{7} =
~~0.124 \pm 0.208$ \\
1.228 & $\eta_{8} = 1.163 \pm 0.019$  & \\
1.070 & $\eta_{9} = 0.972 \pm 0.016$ & \\ \hline
    \hline
  \end{tabular}
\end{table}

As shown in Fig. \ref{mock2p}, all two index parametrization show
deviation from DD relation at $3\sigma$ level. While Fig.
\ref{mock1p}, which shows the likelihood plot for $\eta_{V}$ and
$\eta_{VI}$, again indicates that in case of $\eta_{V}$ there is a
significant deviation from DD relation. Similarly $\eta_{VI}$ only
marginally accommodates the distance duality relation. Thus, the one
index parameterizations also do not support the DD relation
convincingly. It is important to note that five out of six
parameterizations do not give a substantial support to the DD
relation with this bigger mock data set. It may indicate that the
underlying assumptions used to create this mock data set need
revision.

\begin{figure}[ht]
\centering \framebox{
\includegraphics[width=10cm]{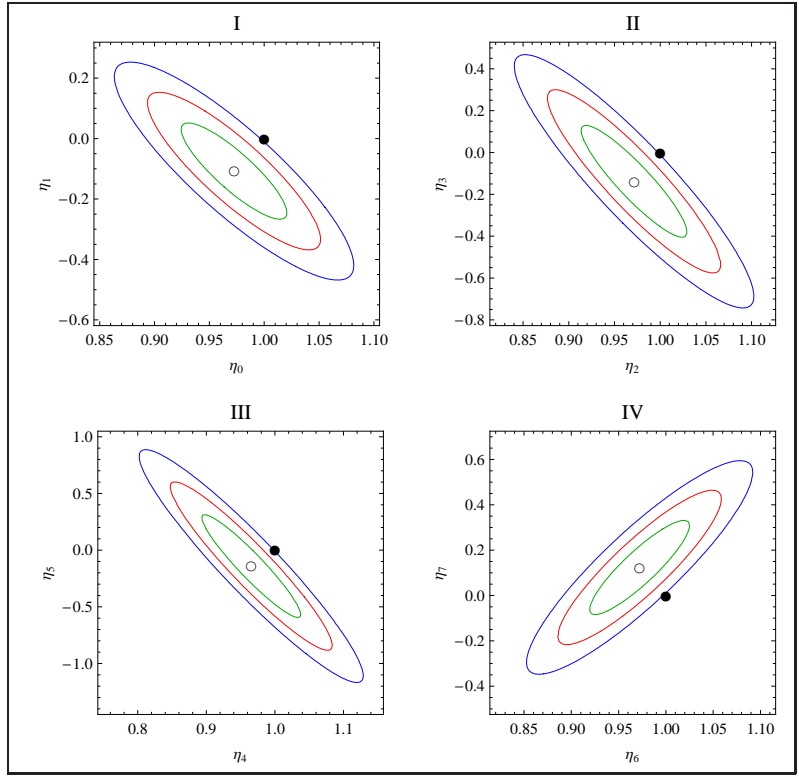}}
\caption{$1 \sigma, 2\sigma$ and $3\sigma$ contours in $\eta_i -
\eta_j$ plane with data set II.  The
  position of filled circle in the contours indicate the point where
  $\eta(z = 0) =
  1$. The position of empty circle indicates the best fit value of the
  parameters.} \label{mock2p}
\end{figure}

\begin{figure}[ht]
\centering \framebox{
\includegraphics[width=8cm]{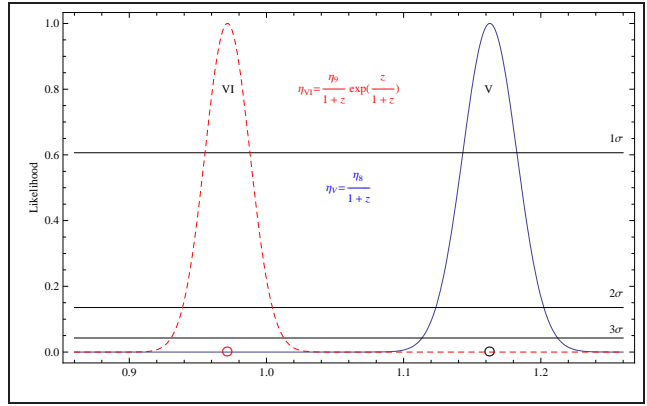}}
\caption{Likelihood distribution from Mock data sample of galaxy
  clusters ( data Set II ). The
position of  empty circle indicate the best fit value of the
parameters.
}\label{mock1p}
\end{figure}

\item In the data set III, the angular diameter distance is obtained
by using the extended physical size of radio galaxies. After using
the selection criterion ($\Delta z <0.005$) we are left with 12 data
points in this set. The best fit values of the parameters
using $\chi^2$ minimum method at $1\sigma$ level are given in Table
\ref{tab:datasetIII}.

\begin{table}[ht]
  \caption{Best fit values for all parameterizations - data set
  III } \vspace{0.2cm}
  \label{tab:datasetIII}
  \centering
  \begin{tabular}{c c c}
    \hline
   {\bf $\chi^2_{\nu}$} & {Parameters} & {Parameters} \\
    \hline
    0.944 & $\eta_{0} = 1.063 \pm 0.198 $ & $\eta_{1} = -0.180 \pm 0.244$  \\
0.971 & $\eta_{2} = 1.099 \pm 0.274 $ & $\eta_{3} = -0.415 \pm 0.632$ \\
1.021 & $\eta_{4} = 1.099 \pm 0.385 $ & $\eta_{5} = -0.749 \pm 1.624$\\
0.958 & $\eta_{6} = 1.081 \pm 0.235 $ & $\eta_{7} =  0.282 \pm 0.404~~$\\
1.592 & $\eta_{8} = 1.564 \pm 0.070 $ &  \\
0.870 & $\eta_{9} = 1.059 \pm 0.047 $ &
\\ \hline
    \hline
  \end{tabular}
\end{table}

As shown in Fig. \ref{daly2pa} and \ref{daly1p}, this data set does
not support DD relation for both 2 index and 1 index
parameterizations convincingly.

\begin{figure}[ht]
\centering \framebox{
\includegraphics[width=10cm]{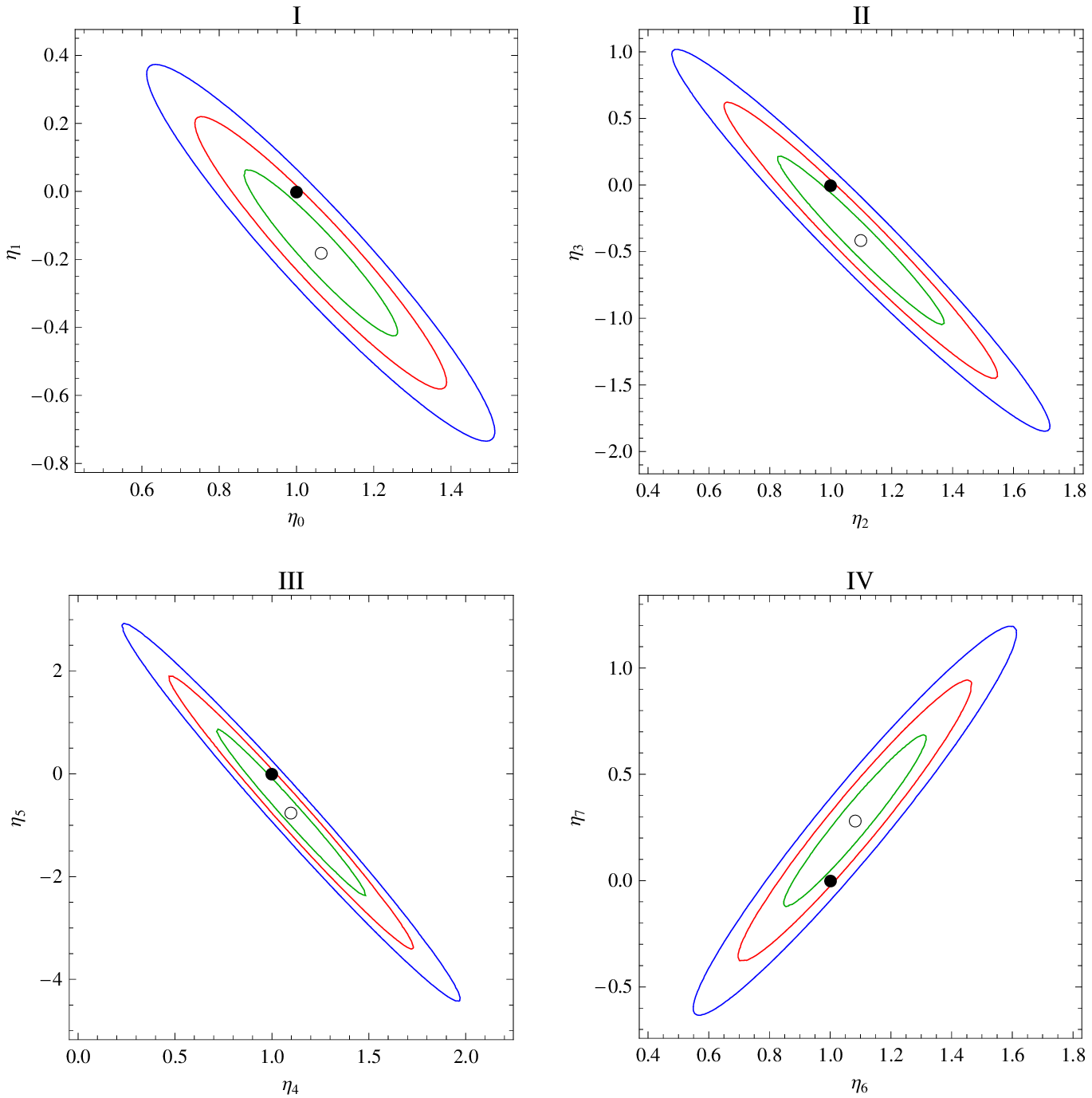}}
\caption{$1 \sigma, 2\sigma$ and $3\sigma$ contours in $\eta_i -
\eta_j$ plane with data set III.  The position of filled circle in
the contours indicate the point where $\eta(z = 0) = 1$. The position of
empty circle indicates the best fit value of the parameters.}
\label{daly2pa}
\end{figure}

\begin{figure}[ht]
\centering \framebox{
\includegraphics[width=8cm]{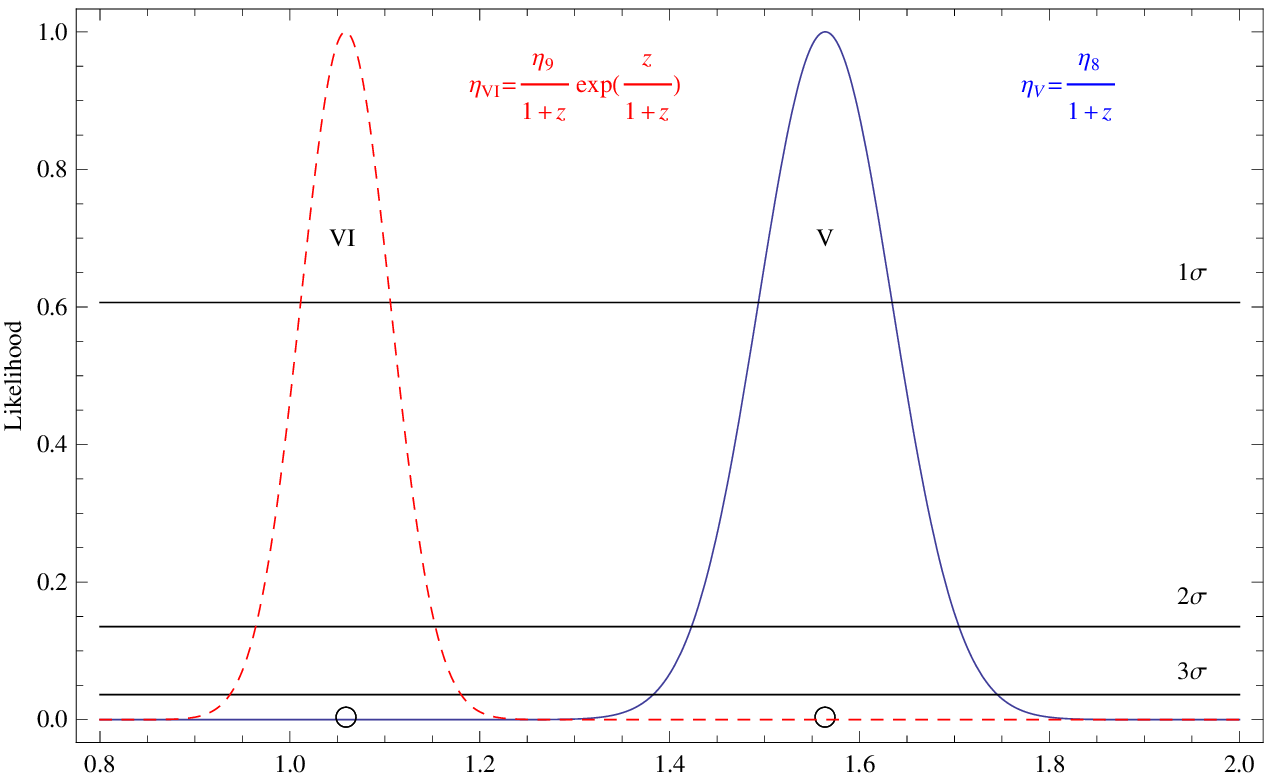}}
\caption{Likelihood distribution function from data set III. The
position of  empty circle indicate the best fit value of the
parameters.} \label{daly1p}
\end{figure}

\begin{figure}
\centering \framebox{
\includegraphics[width=12cm]{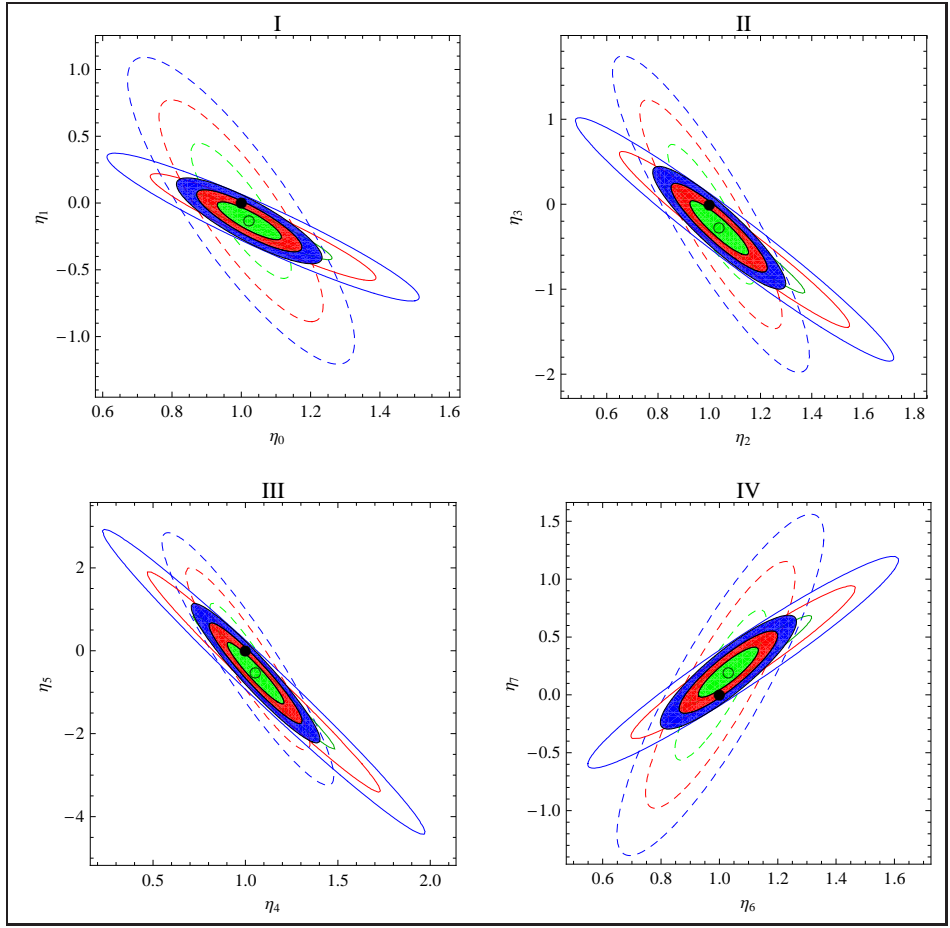}}
\caption{Dashed, solid and filled contours  correspond to data set I, data
set III and  combined data set (I + III )respectively.} \label{combin}
\end{figure}

\item It is important to note that all two index parameterizations
show degenerate behaviour with the given data
sets. These parameterizations are in complete concordance
with DD relation within 1 $\sigma$ level for data set I, shows
deviation at 3 $\sigma$ level for data set II, and in
agreement with DD relation within 2 $\sigma $ level for data set
III. So the behaviour of all two index $\eta(z)$'s strongly depends
upon the data sets chosen here.

\item The $\chi^2_{\nu}$ for $\eta_V$ parametrization with data sets I, II and III is
approximately 1.34, 1.23 and 1.59, respectively. Hence the one index
parametrization, $\eta_{V} $, shows significant deviation with  all
the three data sets. The $ \eta_{VI}$ parametrization marginally
accommodates the DD relation with data set II and III, and
convincingly accommodates data set I. It is evident that for one
index parameterizations the chosen parametrization prior dominates
over the data sets. Therefore the
 one index parameterizations analysed here
are not a good choice for analysing the DD\ relation. This is in
contrast with the behaviour shown by two index parameterizations in
which data seems to dominate. To understand behaviour
of the parameterizations in a qualitative
manner we plot them using their best fit values along with real data
sets (Figs. \ref{fig:parI} and
\ref{fig:parIII}).

\begin{figure}
\centering \framebox{
\includegraphics[width=11cm]{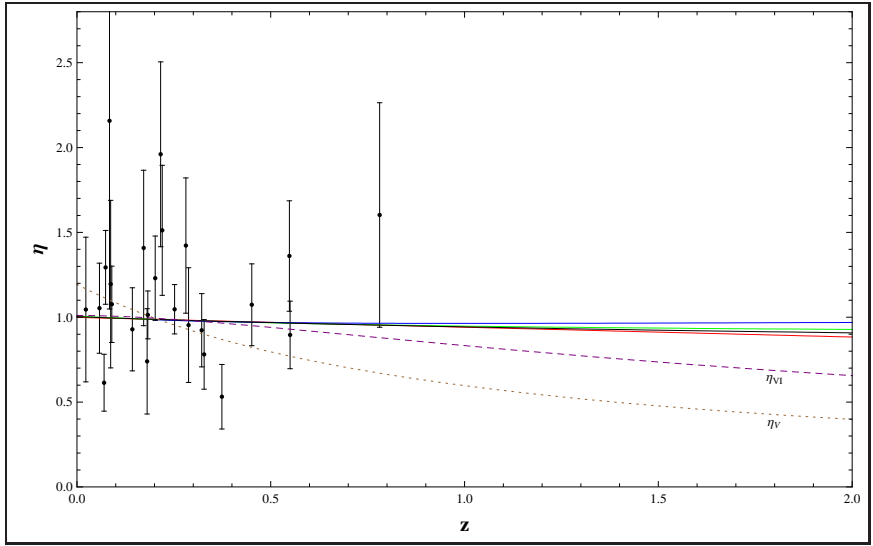}}
\caption{Variation of parameterizations with redshift using best
fit values given in Table 1 . The points with error bars are from data
set I. Solid line
correspond to two
index parameterizations. Dot and dashed lines correspond to one index
parameterizations. }
\label{fig:parI}
\end{figure}

\begin{figure}
\centering \framebox{
\includegraphics[width=11cm]{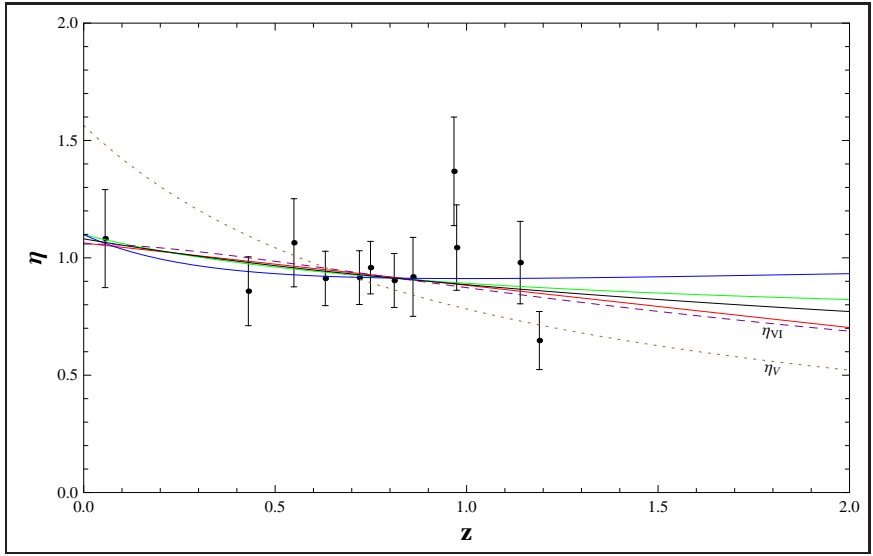}}
\caption{Variation of parameterizations with redshift using best
fit values given in Table 3. The points with error bars are from data set
III. Solid
lines correspond to two index parameterizations. Dot and dashed lines
correspond to one index parameterizations. }
\label{fig:parIII}
\end{figure}

It is clear from the figures \ref{fig:parI} and
\ref{fig:parIII} that parametrization $\eta_{V}(z)$ is
not a good choice for modeling the DD relation. It shows a very
steep deviation from $1$ even at very low redshift ($z \sim 0$),
where we expect DD relation to hold. Hence $\eta_{V}$
parametrization is clearly dominating over the data sets. But on the
other side all the two index parameterizations stay close to unity
(see Figs. \ref{fig:parI} and \ref{fig:parIII}).

\item Out of these three data sets, as expected, the bigger data set of
mock galaxy clusters (data set II) gives tighter constraints on various
parameterizations. But most of the parameterizations show significant deviation from
 DD relation. It is important to note that the mock
 galaxy clusters data
is generated by assuming the spherical isothermal $\beta$  model for clusters.
But the real galaxy cluster data (data set I)  which is obtained by
assuming isothermal
elliptical $\beta$ for clusters is in good agreement with DD relation.

\item The main advantage of using the radio galaxy data is shown
in Fig. \ref{combin} where the contours (dashed lines) correspond to
radio galaxies are aligned to contours corresponding to
clusters (solid line). We  also  show the contours  by using
the combined data set of  real galaxy clusters and  radio galaxies. As expected
the constraints on the parameters are improved.

\item Overall, the pattern seen between one and two-parameter
parameterizations suggest that for the two-parameter case the data
dominates the result, whereas for the one-parameter case the chosen
parameterization prior dominates the result.

\end{enumerate}

When we finished this work, we came across a paper by Li et al. who
also have checked the violation of DD \cite{li}. However, we have
used more data sets, and we have worked with six parameterizations.

\acknowledgments Authors are grateful to anonymous referee for
 useful suggestions. We are grateful to Ruth A.
Daly, S. Khedekar and S. Majumdar for providing the data and  for
useful discussions at various stages of this work. One of the author
(D.J.) also thank A. Mukherjee and S. Mahajan for providing the
facilities to carry out the research. R.N. acknowledges support
under CSIR - JRF scheme (Govt. of India). Authors also acknowledge
the financial support provided by Department of Science and
Technology, Govt. of India under project No. SR/S2/HEP-002/008.

\end{document}